\documentclass[%
 reprint,
superscriptaddress,
showpacs,preprintnumbers,
 amsmath,amssymb,
 aps,
 prb,
floatfix,
]{revtex4-1}
\usepackage[usenames]{color}
\usepackage{graphicx}% Include figure files
\usepackage{dcolumn}% Align table columns on decimal point
\usepackage{bm}% bold math

\newcommand{\CFO}[0]{CoFe$_2$O$_4$}
\newcommand{\NFO}[0]{NiFe$_2$O$_4$}

\begin{document}

\title{Spin-filtering efficiency of ferrimagnetic spinels \CFO\ and \NFO}

\author{Nuala M.~Caffrey} \email{caffreyn@tcd.ie} \altaffiliation[Current
  address: ]{Institute of Theoretical Physics and Astrophysics,
  Christian-Albrechts-Universit\"at zu Kiel, Leibniz\-strasse 15,
  24098 Kiel, Germany} \affiliation{School of Physics and CRANN,
  Trinity College, Dublin 2, Ireland}

\author{Daniel Fritsch} \affiliation{H. H. Wills Physics Laboratory,
  University of Bristol, Tyndall Avenue, Bristol BS8 1TL, United
  Kingdom}

\author{Thomas Archer} \affiliation{School of Physics and
  CRANN, Trinity College, Dublin 2, Ireland} 

\author{Stefano Sanvito} \affiliation{School of Physics and CRANN,
  Trinity College, Dublin 2, Ireland}

\author{Claude Ederer} \email{claude.ederer@mat.ethz.ch}
\affiliation{Materials Theory, ETH Z\"urich, Wolfgang-Pauli-Strasse
  27, 8093 Z\"urich, Switzerland}

\date{\today}

\begin{abstract}
We assess the potential of the ferrimagnetic spinel ferrites \CFO\ and
\NFO\ to act as spin filtering barriers in magnetic tunnel
junctions. Our study is based on the electronic structure calculated
by means of first-principles density functional theory within
different approximations for the exchange correlation energy. We show
that, in agreement with previous calculations, the densities of states
suggest a lower tunneling barrier for minority spin electrons, and
thus a negative spin-filter effect. However, a more detailed analysis
based on the complex band-structure reveals that both signs for the
spin-filtering efficiency are possible, depending on the band
alignment between the electrode and the barrier materials and
depending on the specific wave-function symmetry of the relevant bands
within the electrode.
\end{abstract}

%%%%%\pacs{Valid PACS appear here} \keywords{Suggested keywords}

\maketitle

% General introduction and motivation

\section{Introduction}

The ability to generate and detect spin-polarized currents is a
central requirement for any practical spintronics device. A promising
approach to achieve this goal is to use tunnel junctions containing
ferro- or ferrimagnetic barrier materials, thus presenting different
tunneling probabilities for majority (spin-up, $\uparrow$) and
minority (spin-down, $\downarrow$) electrons. Efficient spin-filtering
has been demonstrated for ferromagnetic insulators such as
EuS,\cite{Moodera_et_al:1988} EuO,\cite{Santos/Moodera:2004} and
BiMnO$_3$.\cite{Gajek_et_al:2005} However the magnetic ordering
temperatures of these magnets are rather low. Therefore the
identification of suitable barrier materials that operate at room
temperature or above is of great interest.

% Summary of experimental results and specific motivation

Spinel ferrites are insulating ferrimagnets with high Curie
temperatures ($T_\text{C}=$790~K for \CFO\ and 865~K for
\NFO),\cite{Brabers:1995} and therefore are promising candidates for
efficient room temperature spin-filtering. A measure of the ability of
a material or a device to select a particular spin direction is the
spin-filtering efficiency, $P_\text{sf}$, which is defined as
$$P_\text{sf}=\frac{I^\uparrow-I^\downarrow}{I^\uparrow+I^\downarrow}\:,$$
where $I^\sigma$ is the spin-$\sigma$ component of the current, which
is assumed to be carried by the two spin species in parallel. Recent
experiments on ferrimagnetic spinels appear promising, as a
spin-filtering efficiency of $+$22\,\% has been measured for \NFO\ at
low temperatures.\cite{Lueders_et_al_APL:2006} The measured positive
sign of $P_\text{sf}$ is in apparent contradiction with results of
band-structure calculations, demonstrating that the bottom of the
conduction band is lower for spin-down electrons than for
spin-up,\cite{Szotek_et_al:2006} which would lead to a lower tunneling
barrier for minority spin electrons.  It was suggested that this
apparent discrepancy could be due to effects related to the
wave-function symmetry of the tunneling
states.\cite{Lueders_et_al_APL:2006} Furthermore, for \CFO\ both
positive and negative $P_\text{sf}$ have been reported in junctions
made of different electrode materials and where $P_\text{sf}$ was
measured with different experimental techniques. The reported values
of $P_\text{sf}$ range from $-$44\,\% to
$+$26\,\%.\cite{Chapline/Wang:2006,Chen/Ziese:2007,Ramos_et_al_APL:2007,Ramos_et_al:2008,Rigato_et_al:2010,Takahashi_et_al:2010}
Due to these large variations in experimental results (with both signs
occurring for the spin-filtering efficiency) a conclusive picture of
spin-filtering in spinel ferrites has not emerged, yet. As such, a
first-principles investigation of the spin-filtering efficiency in
these materials is highly desirable, in order to provide a reference
for future experimental studies and to allow further optimization of
the corresponding devices.

% State of the art (and its limitations) of current spin filter theory
% for spinel ferrites

So far, theoretical predictions for the spin-filter effect in
\CFO\ and \NFO\ are almost exclusively based on density of states
(DOS) calculations within a self-interaction corrected (SIC) local
spin-density approximation (LSDA).\cite{Szotek_et_al:2006} The
spin-splitting of the conduction band minimum (CBM) in these
calculations suggests a lower tunnel barrier for minority spin
electrons and thus a negative sign for the spin-filtering
efficiency. However, it is well known that in many cases this simple
density of states argument can be misleading, and the tunnel
probability can be strongly dependent on the specific wave-function
symmetry.\cite{Mavropoulos_PRL85_1088} The implications of this were
first noticed in a Fe/MgO/Fe
heterostructure,\cite{PhysRevB.63.054416,PhysRevB.63.220403} where
symmetry-dependent tunneling results in half-metallic behaviour of the
Fe/MgO(001) stack. Since then, the so-called \emph{complex
  band-structure}, which determines the decay length of Bloch states
with different wave-function symmetries inside an insulating barrier,
has been used to account for many, otherwise unexplained, experimental
results in spin-dependent tunnel junctions.  Furthermore, it is of
interest to compare the SIC-LSDA result of
Ref.~\onlinecite{Szotek_et_al:2006} to the electronic structure
obtained by using alternative approaches such as LSDA+$U$, hybrid
functionals, or other SIC approaches.

% Preview of paper and summary of main results.

Here we present a detailed comparison of the electronic structure of
\CFO\ and \NFO\ calculated within different approximations for the
exchange-correlation potential. This allows us to identify features of
the DOS that are fairly robust with respect to the specific choice of
exchange-correlation potential and features that are very sensitive to
this choice.  In addition, we calculate the complex band-structure for
both materials within the atomic SIC method
(ASIC)\cite{Pemmaraju/Sanvito:2007,VASIC}, which facilitates the
identification of suitable electrode materials that can lead to high
spin-filtering efficiency. We show that, for both \CFO\ and \NFO\ and
the two transport directions [001] and [111], electrons tunnel with
the highest probability at the center of the two-dimensional Brillouin
zone in the plane orthogonal to the transport direction. Furthermore,
depending on the exact alignment of the electrode Fermi level relative
to the CBM of the barrier, the tunneling current may present either a
predominant majority or a predominant minority contribution,
i.e. $P_\text{sf}$ may change sign depending on the level alignment.

The paper is organized as follows. After having briefly presented the
computational method and the details of the crystallographic unit cell
used for this study, we proceed to describe the electronic structure
of \CFO\ and \NFO. In particular, we first discuss the DOS and real
band-structures, and then move on to present the complex ones. The
final section summarizes our main conclusions.

% Computational details

\section{Methods}

We employ the {\sc vasp}\cite{Kresse_CompMatSci6_15} and {\sc
  siesta}\cite{0953-8984-14-11-302} density functional theory (DFT)
code packages for the calculation of DOS and real band-structures and
the {\sc smeagol} code\cite{PhysRevB.73.085414, Smeagol2} to calculate
the complex band-structure. The {\sc vasp} calculations have been
performed by using the projector-augmented wave (PAW)
method\cite{Bloechl_PRB50_17953} with standard PAW potentials supplied
with the {\sc vasp} distribution, a 500~eV plane wave energy cutoff,
and a $\Gamma$ centered 6$\times$6$\times$6 $k$-point mesh for the
Brillouin zone sampling. We employ the generalized gradient
approximation (GGA) according to the Perdew-Burke-Ernzerhof
formulation \cite{Perdew/Burke/Ernzerhof:1996} together with the
Hubbard ``$+U$'' correction,\cite{Anisimov_JPCM9_767} where $U = 3$ eV
and $J = 0$ eV is applied to the $d$ states of all transition metal
cations, as well as the hybrid functional approach according to Heyd,
Scuseria and Ernzerhof (HSE),\cite{Heyd/Scuseria/Ernzerhof:2003} using
the standard choice for the fraction of Hartree-Fock exchange
($\alpha=0.25$) and a reduced plane wave energy cutoff of
400\,eV. When using the localised basis set code {\sc siesta},
structural relaxations were performed using the GGA while the atomic
self-interaction correction (ASIC) scheme was used to determine the
electronic structure, including the complex band-structure.  A
6$\times$6$\times$6 $k$-point Monkhorst-Pack mesh was used to converge
the density matrix to a tolerance of $10^{-5}$ and a grid spacing
equivalent to a plane-wave cutoff of 800\,eV was used.

% Spinel specific details (structure)

For most of our calculations we use the smallest possible unit cell
(containing 2 formula units) to describe the inverse spinel
structure. The corresponding distribution of cations on the spinel $B$
site lowers the space group symmetry from $Fd\bar{3}m$ to
$Imma$.\cite{Fritsch/Ederer:2010} We also present some results
obtained for a cation distribution with $P4_122$ symmetry, which
requires a doubling of the unit cell to 4 formula units (the $k$-point
sampling is then adjusted accordingly). We have previously shown that
both $Imma$ and $P4_122$ are low energy configurations for the inverse
spinel structure in \CFO\ and \NFO, and that the specific cation
arrangement has only a minor influence on the global electronic
structure of these systems.\cite{Fritsch/Ederer:2011} We note that
experimentally a disordered distribution of Fe$^{3+}$ and
Co$^{2+}$/Ni$^{2+}$ cations over the spinel $B$ site with effective
cubic $Fd\bar{3}m$ symmetry, i.e. with no long-range cation order, is
generally observed, even though recently indications for short range
cation order in both \NFO\ bulk and thin film samples have been
reported.\cite{Ivanov_et_al:2010,Iliev_et_al:2011} For a more detailed
comparison between the different cation configurations see
Refs.~\onlinecite{Fritsch/Ederer:2011} and
\onlinecite{Fritsch/Ederer:2012}.

Structural relaxations have been performed at the GGA level, with all
cations being fixed to their ideal cubic
positions.\cite{Fritsch/Ederer:2010} The relaxed bulk lattice
constants $a_0$ obtained by using {\sc vasp} ({\sc siesta}) are 8.366
\AA{} (8.360 \AA{}) and 8.346 \AA{} (8.356 \AA{}) for \CFO\ and \NFO,
respectively, and are in very good agreement with experimental data
(see Ref.~\onlinecite{Fritsch/Ederer:2011} and references therein).

\section{Results and Discussion}

\subsection{Electronic structure}

\begin{figure*}
\begin{center}
\includegraphics[width=\textwidth,clip]{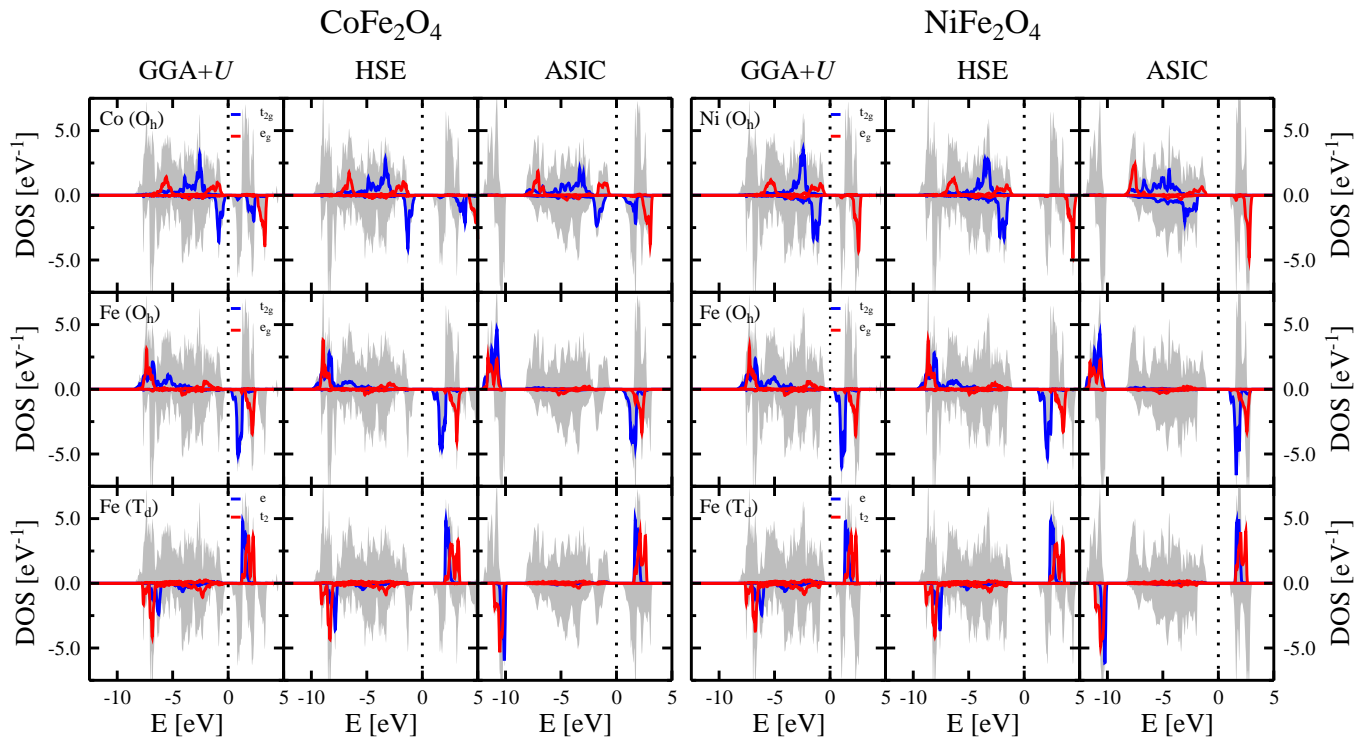}
\caption{\label{Fig_DOS}(Color online) Total and projected DOS per
  formula unit for \CFO\ (left panels) and \NFO\ (right panels)
  calculated with different exchange-correlation potentials (from left
  to right: GGA$+U$, HSE and ASIC). The $t_{2g}$ and $e_g$ states of
  Fe, Co, and Ni on the $O_h$ sites and the $e$ and $t_2$ states of Fe
  on the $T_d$ sites are shown as black (blue) and dark grey (red)
  lines, respectively. The shaded grey area in all panels depicts the
  total DOS. Minority spin projections are shown using negative
  values. The zero energy is set to the middle of the band gap.}
\end{center}
\end{figure*}

\begin{table}
\caption{\label{TableElectronicProperties}Band gap (E$_g$) and
  spin-splitting of the CBM ($\Delta$CBM) for \CFO\ and
  \NFO\ calculated with different exchange-correlation
  functionals. All values are in eV.}
\begin{ruledtabular}
\begin{tabular}{l|cc|cc}
& \multicolumn{2}{c|}{\CFO} & \multicolumn{2}{c}{\NFO} \\ 
& $E_g$ & $\Delta$CBM & $E_g$  & $\Delta$CBM \\
\hline 
GGA$+U$ & 0.52 & 0.92 & 0.83 & 0.86 \\ 
HSE     & 1.60 & 1.09 & 2.32 & 1.00 \\ 
ASIC    & 1.08 & 1.00 & 2.07 & 0.46 \\
\end{tabular}
\end{ruledtabular}
\end{table}

It has been previously shown that GGA leads to a half-metallic
solution for \CFO\ and results in only a very small insulating gap in
the case of \NFO\ (see e.g. Refs.~\onlinecite{Fritsch/Ederer:2010} and
\onlinecite{Fritsch_JPhysConfSer292_012104} and references
therein). The DOS of \CFO\ and \NFO\ calculated by using a selection
of beyond-GGA functionals are depicted in Fig.~\ref{Fig_DOS}. It can
be seen that all the studied exchange-correlation potentials lead to
an insulating state for \CFO\ and an enhanced band gap for \NFO. When
compared to the GGA+$U$ band gaps, both the inclusion of Hartree-Fock
exchange within the HSE calculation as well as the ASIC treatment
leads to a large increase in the band gap values for both the Co and
Ni based ferrite, with the largest band gaps obtained for HSE (see
Table~\ref{TableElectronicProperties}). We also note that our results
are consistent with recent HSE and LSDA+$U$ calculations for
\NFO.\cite{Sun_et_al:2012}

Going into more details we notice that, while the occupied DOS are
very similar for GGA+$U$ and HSE, the ASIC method places the local Fe
spin-majority states significantly lower in energy. This results in a
gap between these Fe states and the higher-lying Co (Ni) $d$ and
oxygen $p$ valence bands. Interestingly, for \CFO\ the valence band
maximum in ASIC is made up of the majority spin Co $e_g$ states,
whereas for both GGA+$U$ and HSE the corresponding minority spin
$t_{2g}$ states are slightly higher in energy.  We also note that the
difference in the calculated GGA+$U$ band gap of \CFO\ (\NFO) compared
to the previously obtained values of 0.9~eV (0.97~eV) for the $Imma$
structure,\cite{Fritsch/Ederer:2010} and 1.24~eV (1.26~eV) for the
$P4_{1}22$ structure\cite{Fritsch/Ederer:2011}, is due to the fact
that in the present work all calculations are performed at the GGA
volume, whereas the calculations in
Refs.~\onlinecite{Fritsch/Ederer:2010} and
\onlinecite{Fritsch/Ederer:2011} have been performed at the larger
GGA+$U$ optimized volume. In addition to the expected dependence on
the exchange correlation potential, our results thus indicate a strong
volume sensitivity in particular of the calculated \CFO\ band
gap. Experimental estimates for the band gaps of spinel ferrites are
sparse and vary over a broad range comprised between 0.11\,eV and
1.5\,eV for \CFO\ and between 0.3\,eV and 3.7\,eV for
\NFO.\cite{Waldron:1955,Rai_et_al:2011} A recent optical absorption
study of \NFO\ suggests an indirect gap of 1.6\,eV in the minority
spin channel,\cite{Sun_et_al:2012} which thus represents an upper
bound for the corresponding fundamental band gap.

In all cases, and for both \CFO\ and \NFO, the CBM is lower in energy
for the spin-down states than for spin-up ones, in agreement with the
SIC-LSDA calculations of Ref.~\onlinecite{Szotek_et_al:2006}. In the
case of \CFO, all the three approaches used in our work predict a
spin-splitting of the CBM ($\Delta$CBM in
Table~\ref{TableElectronicProperties}) of around 1~eV. For
\NFO\ however, GGA+$U$ and HSE yield a $\Delta$CBM of around
0.9-1.0~eV, while ASIC gives a somewhat smaller splitting of only
0.46~eV. In all the cases, the obtained spin-splittings of the CBM are
smaller than those reported in Ref.~\onlinecite{Szotek_et_al:2006},
1.28~eV (1.21~eV) for \CFO\ (\NFO). We note, however, that even
smaller values, namely of 0.47~eV for both \CFO\ and \NFO, have been
obtained in previous GGA+$U$ calculations at the relaxed GGA+$U$
volume.\cite{Fritsch/Ederer:2011} Recent experiments estimate the
spin-splitting of the CBM in the tens of meV range for \CFO-containing
junctions.\cite{Ramos_et_al_APL:2007} 

\begin{figure}
\begin{center}
\includegraphics[width=\columnwidth,clip]{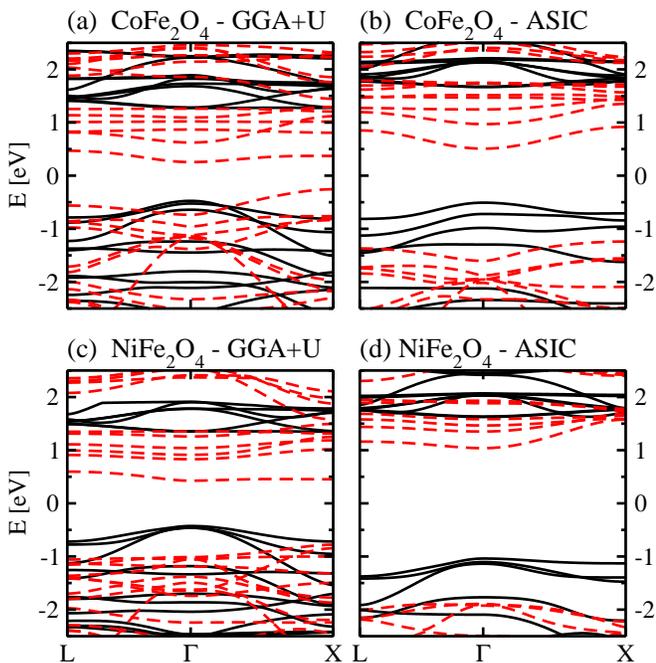}
\caption{\label{Fig_BAND}(Color online) Band structures for energies
  around the band gap of \CFO\ [upper panels (a) and (b)] and
  \NFO\ [lower panels (c) and (d)] calculated by using the GGA$+U$
  exchange-correlation functional [left panels (a) and (c)] and the
  ASIC scheme [right panels (b) and (d)]. Majority and minority spin
  bands are shown as full (black) and dashed (red) lines.}
\end{center}
\end{figure}

In order to shed further light on the nature of the bands around the
gap, the calculated GGA+$U$ and ASIC band-structures for both
\CFO\ and \NFO\ are shown in Fig.~\ref{Fig_BAND}. Apart from the
larger band gaps obtained by the ASIC approach, it can be seen that
the relative energies of the minority and majority spin bands in the
upper valence band region for \CFO\ differ between GGA+$U$ and
ASIC. This is consistent with our previous discussion of the DOS. For
the calculation that is performed with GGA+$U$, the top of the valence
band is formed by a minority spin band with maximum at the X point,
i.e. the minority spin band gap is indirect. In contrast a direct gap
with mixed spin character at $\Gamma$ is obtained by ASIC. Since, as
we will show in the following, the tunneling probabilities are
dominated by states around the $\Gamma$ point, we do not expect that
this qualitative difference between the two exchange-correlation
functionals will critically affect the transport properties.

Based on our analysis of the DOS and the band-structure in the
vicinity the gap, we can conclude that despite some differences, all
computational methods consistently predict a lower tunnel barrier for
the minority spin electrons and therefore a negative spin-filtering
efficiency for both \CFO\ and \NFO. However, as shown in Fe/MgO/Fe
tunnel junctions, \cite{PhysRevB.63.054416,PhysRevB.63.220403} in the
case of high quality epitaxial interfaces between the electrodes and
the barrier material such DOS considerations are only of limited value
for the description of actual transport properties. Instead, the
specific symmetry of the decaying wave-functions inside the barrier
has to be considered. This can be achieved through calculation of the
complex band-structure.\cite{Mavropoulos_PRL85_1088}

%%%%%%%%%%%%%%%%%%%%%%%%%%%%%%%%%%%%%%%%%%%%%%%%%%%%%
% Complex band structure: begin theoretical background

\subsection{Complex band structure}

The complex band-structure along a particular crystalline direction is
calculated with the DFT non-equilibrium Green's function code {\sc
  smeagol}.\cite{PhysRevB.73.085414, Smeagol2} The complex
band-structure is nothing but the solution of the secular band
equation extended to imaginary wave-vectors. Let us assume that the
transport direction of a given tunnel junction is along the $z$
direction and that the material composing the barrier has a particular
crystalline axis aligned along that direction. For any given
$k$-vector in the transverse $x$-$y$ plane, $k_\parallel=(k_x,k_y)$,
and for any energy, $E$, the band equation $E=E(k_x,k_y,k_z)$ can be
solved for $k_z$ if one admits imaginary solutions
$k_z=q+i\kappa$. This means that the wave-function of an electron
approaching the tunneling barrier with transverse wave-vector
$k_\parallel$ exponentially decays into the barrier along the $z$
direction over a length-scale given by $1/\kappa$. Clearly such decay
rate depends on the transverse $k$-vector and the energy, i.e.
$\kappa=\kappa(k_x,k_y;E)$. Here we consider the situation of electron
transport along both the [001] and [111] directions of the cubic
spinel structure.

% Complex band structure: end theoretical background
%%%%%%%%%%%%%%%%%%%%%%%%%%%%%%%%%%%%%%%%%%%%%%%%%%%%%

\begin{figure}
\begin{center}
(a) $Imma$-\CFO, transport along
  [001]\\ \includegraphics[width=0.9\columnwidth,clip]{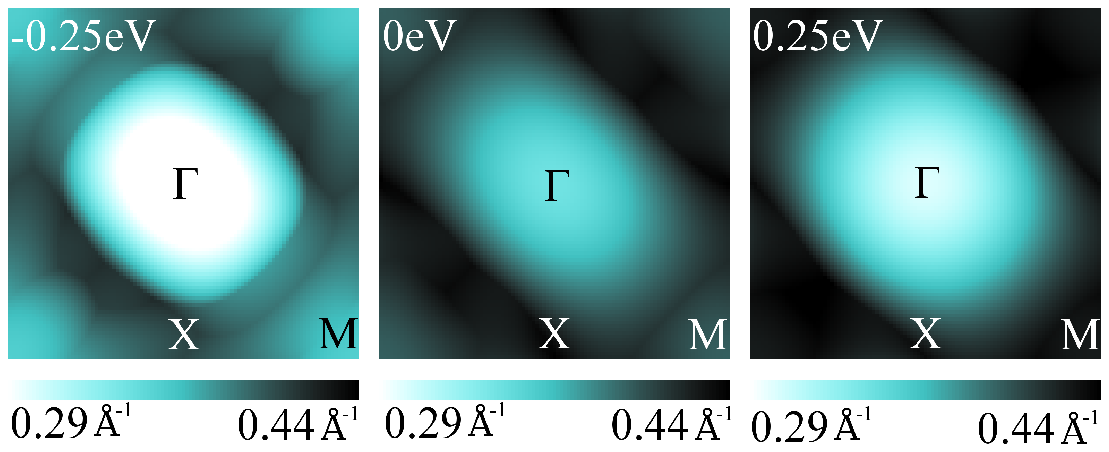}\\ (b)
  $Imma$-\CFO, transport along
  [111]\\ \includegraphics[width=0.9\columnwidth,clip]{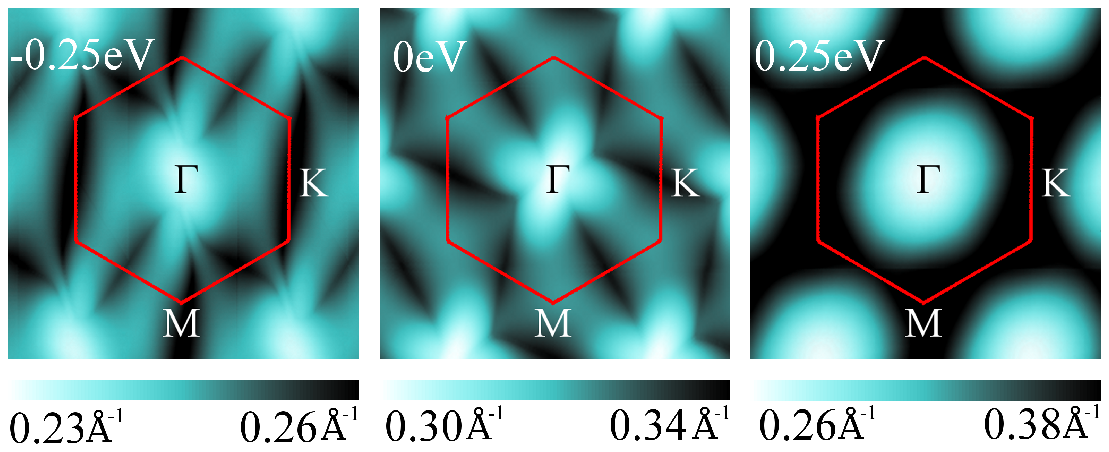}\\ (c)
  $P4_122$-\CFO, transport along
  [001]\\ \includegraphics[width=0.9\columnwidth,clip]{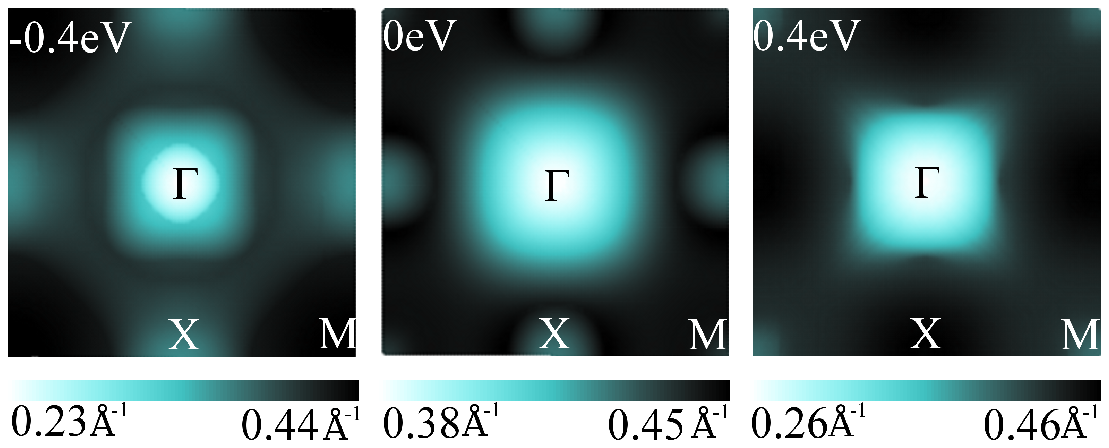}\\ (d)
  $Imma$-\NFO, transport along
  [001]\\ \includegraphics[width=0.9\columnwidth,clip]{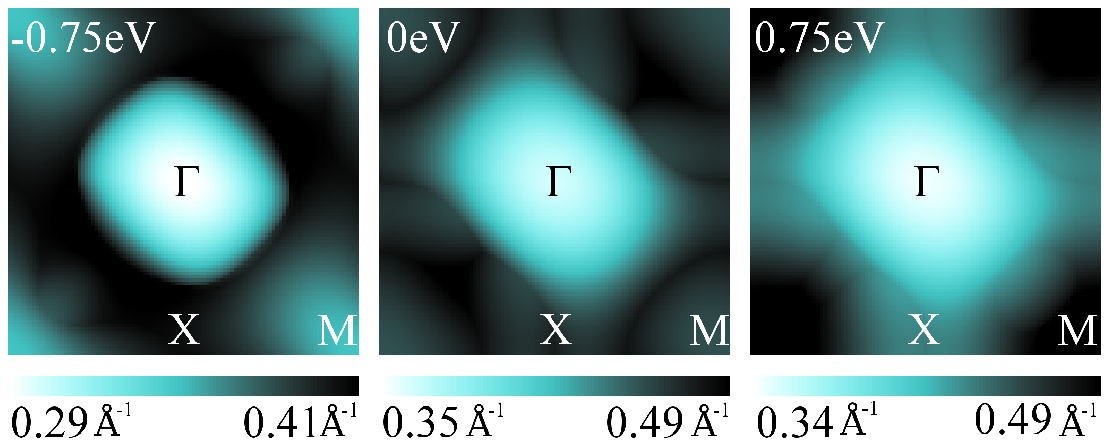}\\ (e)
  $Imma$-\NFO, transport along
  [111]\\ \includegraphics[width=0.9\columnwidth,clip]{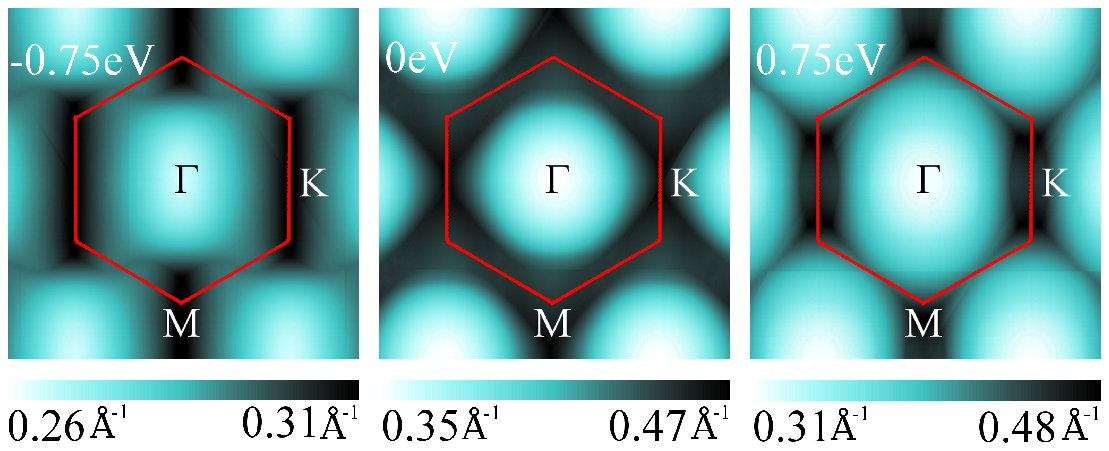}\\ (f)
  $P4_122$-\NFO, transport along
  [001]\\ \includegraphics[width=0.9\columnwidth,clip]{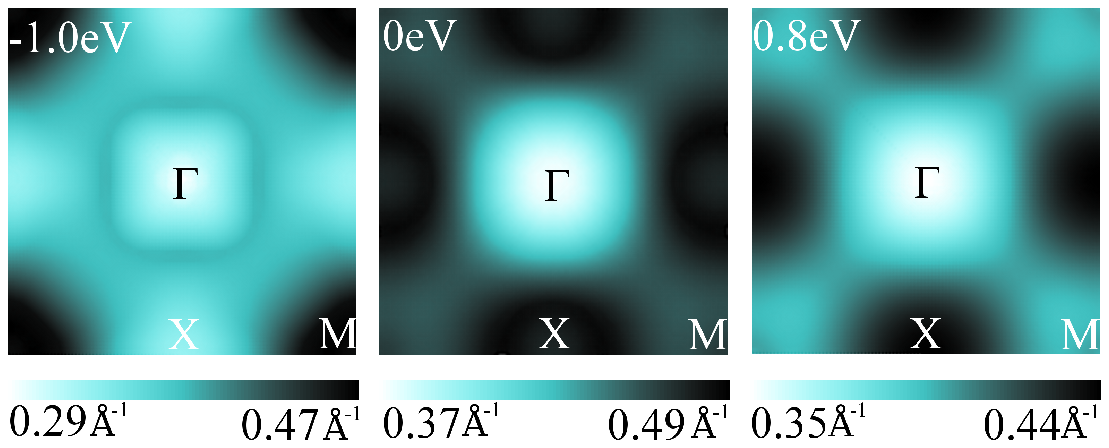}
\caption{\label{fig:kxky} Minimal value of $\kappa$ at different
  energies (indicated at the top left in each graph) within the gap
  for \CFO\ (a-c) and \NFO\ (d-f) along different transport
  directions, calculated within the ASIC approach. Zero energy
  corresponds to the middle of the band gap.}
\end{center}
\end{figure}

In Fig.~\ref{fig:kxky} we plot the minimal value of $\kappa$ as a
function of $k_x$ and $k_y$ (calculated on a 100$\times$100 grid) at
different energies within the gap. We include data for both \CFO\ and
\NFO\ considering both transport directions for the $Imma$
configuration, and we also present data for the $P4_122$ configuration
and transport along the [001] direction. The crucial result emerging
from Fig.~\ref{fig:kxky} is that in all cases $\kappa$ is smallest at
the $\Gamma$ point of the two-dimensional Brillouin zone corresponding
to the $x$-$y$ plane. This means that, due to the exponential
dependence of the wave-function on $\kappa$, electron tunneling away
from the $\Gamma$ point will contribute very little to the
transport. As such, in the analysis that will follow, we will only
consider transport through the $\Gamma$-point. We note that
$\Gamma$-point filtering is a highly desirable property for both
tunnel junctions and spin injection. As has been shown for the Fe/MgO
barrier, as the thickness of the MgO layer increases so does the
selectivity of the $\Gamma$-point. This in turn increases the
tunneling magneto-resistance (TMR). Although the $\Gamma$-point
filtering is not strictly necessary for a large TMR, it significantly
reduces the importance of the material choice for the electrodes.

\begin{figure}
\begin{center}
\includegraphics[width=\columnwidth,clip]{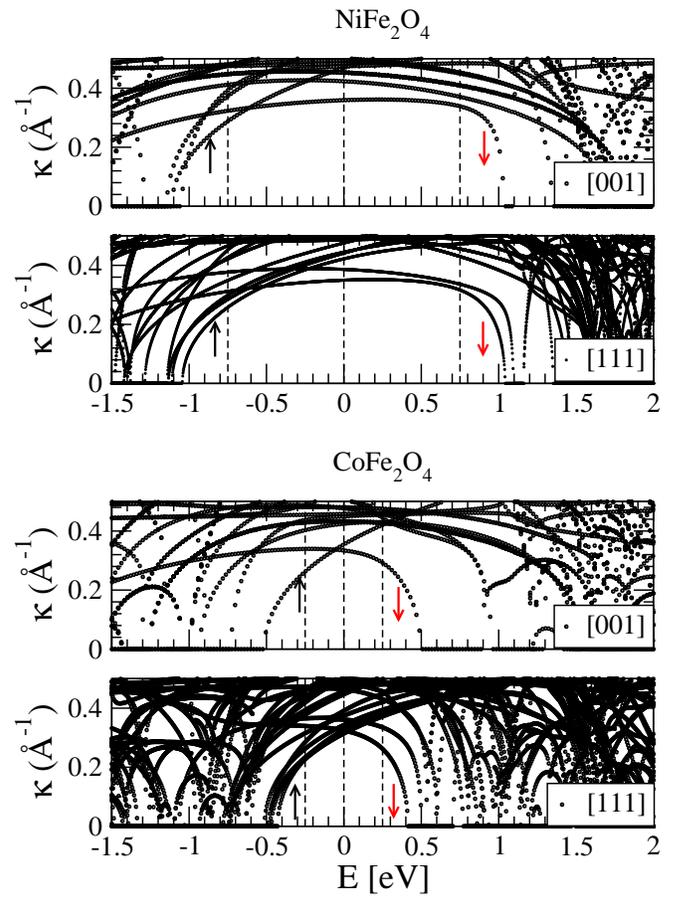}
\caption{\label{fig:cbs_combined} The complex band-structure
  corresponding to $k_x=k_y=0$ for \NFO\ (upper two panels) and
  \CFO\ (lower two panels) along [001] and [111], calculated within
  ASIC for the $Imma$ ionic configuration. The up and down arrows
  indicate the spin-character of the lowest lying complex bands. The
  vertical dashed lines indicate the energies that were used for the
  $k_x$-$k_y$ plots in Fig.~\ref{fig:kxky}.}
\end{center}
\end{figure}

Having established that the transport predominantly occurs at the
$\Gamma$-point, further insight can be gained by exploring the energy
dependence of $\kappa(0,0;E)$. In particular it is important to
establish the spin and orbital symmetry of the complex bands
corresponding to the smallest value of $\kappa(0,0;E)$ for each
energy, since incident waves with that particular symmetry will
dominate the tunneling current. In Fig.~\ref{fig:cbs_combined} we
present the complex band-structures of \CFO\ and \NFO, calculated
along the [001] and [111] directions at the $\Gamma$-point in the
transverse 2D Brillouin zone for the $Imma$ configuration. One can
easily recognize that, for both \CFO\ and \NFO, the main features
which we discuss in the following are similar for the two different
transport directions. We note that the transport calculation along
[111] requires a larger unit cell, in order to obtain lattice vectors
that are either perpendicular or parallel to the transport direction,
which leads to a larger number of complex bands compared to the [001]
case. In both materials the slowest decay rate close to the valence
band maximum corresponds to electrons with majority spin character (in
agreement with the real band-structure shown in
Fig.~\ref{Fig_BAND}). This remains the case for energies up to around
0.5~eV from the top of the valence band, although the decay rate
increases quickly with energy. In contrast, the lowest decay rate for
energies taken in the upper part of the band gap is dominated by
states with minority spin symmetry. For \NFO\ this decay rate remains
almost constant for a wide energy window of about 1.5~eV, whereas for
\CFO\ the gap region is divided more symmetrically between the
majority and minority spin-dominated regions. The smaller ASIC
calculated band gap of \CFO\ compared to that of \NFO\ results in
slightly slower decays within the gap region for both majority and
minority spins.

\begin{figure}
\begin{center}
\includegraphics[width=\columnwidth,clip]{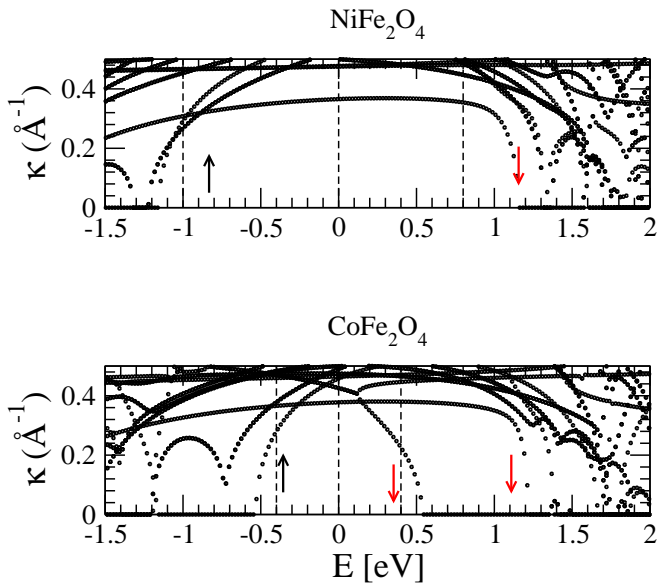}
\caption{\label{fig:cbs_p4122} The complex band-structure
  corresponding to $k_x=k_y=0$ for \NFO\ (upper panel) and
  \CFO\ (lower panel) along [001] for the $P4_122$ configuration,
  calculated within ASIC. The up and down arrows indicate the
  spin-character for some of the lowest lying complex
  bands.}
\end{center}
\end{figure}

In Fig.~\ref{fig:cbs_p4122} we also present the complex band structure
of \CFO\ and \NFO\ in the $P4_122$ configuration for transport along
[001]. One can recognize the slightly larger band-gap compared to the
$Imma$ configuration, but for \NFO\ the complex bands look very
similar compared to the $Imma$ case. For \CFO\ one can see that the
bands in the mid-gap region connect in a somewhat different way than
in the $Imma$ configuration. However, the spin-characters of the
lowest complex band in the upper and lower gap region remain
unaffected by the different cation distribution, even though the
energy range dominated by the minority spin complex bands is somewhat
more extended in the $P4_122$ case.

From the complex bands it becomes clear that positive as well as
negative values for $P_\text{sf}$ are possible for both \NFO\ and
\CFO, depending on whether the Fermi level of the electrode lies in
the upper or lower gap region of the spinel tunnel barrier, and on the
availability of majority or minority spin carriers in the metal. If
the Fermi level of the metallic electrode lies within $\sim$0.5~eV
from the top of the valence band, the slowest decay rate in both
\CFO\ and \NFO\ will be for electrons with majority spin. In contrast,
if the Fermi level of the electrode is more than 0.5~eV above the
valence band edge of the spinel barrier, then the slowest decaying
state is in the minority spin channel. The exact position of the Fermi
level of the metal depends on the band alignment between the two
materials. Thus, the spin filter efficiency of the spinel ferrite
barrier will depend strongly on the band alignment and eventually also
on the orbital symmetry of the electrode states at the Fermi level. In
addition, a good lattice match is of course required, otherwise
translational symmetry is broken in the transverse plane and the
complex band-structure argument breaks down. Here, the possibility to
grow good quality films of \CFO\ and \NFO\ with either [001] or [111]
orientation (see e.g. Refs~\onlinecite{Chinnasamy_et_al:2007},
\onlinecite{Ramos_et_al_PRB:2007}, and \onlinecite{Ma_et_al:2010})
opens up a wide range of possible electrode materials. In fact, high
quality epitaxial junctions of \CFO\ or \NFO\ with various electrode
materials, such as La$_{2/3}$Sr$_{1/3}$MnO$_3$, Au, Fe$_3$O$_4$,
Nb-doped SrTiO$_3$, Pt, Co, Al, and SrRuO$_3$, have already been
fabricated.\cite{Lueders_et_al_APL:2006,Chapline/Wang:2006,Chen/Ziese:2007,Ramos_et_al_APL:2007,Ramos_et_al:2008,Rigato_et_al:2010,Takahashi_et_al:2010}

So far we have only discussed the spin character of the complex bands,
whereas it is well known from the Fe/MgO/Fe system, that the orbital
character of the relevant bands can also have a crucial influence on
the tunneling properties. The determination of orbital character of
the complex bands in the inverse spinel ferrites \CFO\ and \NFO\ is
complicated by the different symmetries of the specific cation
configurations used in the calculations. For example, the lowest lying
state above the gap at $\Gamma$ in $Imma$-\NFO, i.e. the one which
connects to the complex band with minority spin character that has the
smallest extinction coefficient over a rather large energy region
within the gap, transforms according to the fully symmetric
irreducible representation $A_g$ of the corresponding orthorhombic
point group $mmm$. This means that, assuming an electrode with cubic
bulk symmetry, this state can in principle couple to $\Delta_1$ and
$\Delta_2/\Delta'_2$ bands for transport along the [001] direction
(whether $\Delta_2$ or $\Delta'_2$ depends on how exactly the
electrode is oriented with respect to the spinel structure), or to
$\Lambda_1$ and $\Lambda_3$ for transport along the [111]
direction. However, these consideration hold only for the case with
$Imma$ symmetry and it is unclear how different cation arrangement, in
particular a completely disordered cation distribution, would change
these symmetry-based selection rules. Generally, the lower symmetry of
the various cation arrangements leads to fewer symmetry restrictions
regarding the possible coupling with electrode bands. Since a full
symmetry analysis of all combinations that can possibly occur is
beyond the scope of this paper, we restrict our analysis to the spin
character of the decaying states within the barrier, which was
discussed in the preceeding paragraphs.

\section{Summary and conclusions}

In summary, we have calculated the electronic structure of both
\NFO\ and \CFO\ using different approaches to evaluate the
exchange-correlation potential. These include GGA, GGA+$U$, HSE and
ASIC. We found that, while there are certain characteristic
differences in the predicted band-structure, the densities of states
of all beyond-GGA methods consistently suggest a lower tunnel barrier
for minority spin electrons. Due to the well-known limitations of this
simple density of states picture of tunneling, we have further
analyzed the complex bands of the two materials at the ASIC level. 

We have shown that the tunneling along the [001] and [111] directions
is dominated by zone-center contributions ($k_x=k_y=0$), and that for
both \NFO\ and \CFO\ the spin character of the slowest decaying state
changes within the gap. Therefore, \NFO\ and \CFO\ are both capable of
acting as either positive or negative spin filters, depending on the
band alignment and wave-function symmetry of the electrodes. Given
such a relatively sensitive dependence of the tunneling current on the
position of the electrode Fermi level, we envision that gating may
allow the spin filtering to be switched from positive to negative.

However, we also want to note that based on the complex band-structure
of the barrier alone, it is not possible to make a definite prediction
about the transport properties observed in a specific experiment. One
may still encounter a situation where incident wave-functions with the
desired symmetry, i.e. matching that of the smallest $\kappa(0,0;E)$
inside the barrier, are not available within the electrodes, simply
because of the corresponding real
band-structure\cite{Nuala2011,AdvMat2012}. Furthermore, it has been
demonstrated recently for the case of an Fe-MgAl$_2$O$_4$-Fe tunnel
junction, i.e. containing a non-magnetic spinel as barrier material,
that the different unit cell sizes of the spinel barrier and the Fe
electrodes can open up new transport channels due to ``backfolding''
of bands from the in-plane Brillouin zone boundary onto the $\Gamma$
point.\cite{Miura_et_al:2012} This leads to a relatively low tunnel
magneto-resistance for the Fe-MgAl$_2$O$_4$-Fe junction, even though
the corresponding complex and real band structures would indicate a
highly symmetry-selective
barrier.\cite{Zhang/Zhang/Han:2012,Miura_et_al:2012} Therefore, in
order to fully assess the spin-filter efficiency for a specific
combination of electrode and barrier materials, a full transport
calculation for the entire device needs to be performed. Nevertheless,
the analysis of the complex band-structure provides a powerful
interpretative tool and offers a good indication on what are the
dominant contributions to the tunneling current. In the present case,
it allows the rationalization of both signs of the spin-filter
efficiency occurring in \NFO\ and \CFO\ tunnel junction, depending on
the band alignment with the electrode.

\begin{acknowledgments}
This work has been supported by Science Foundation Ireland under
(Grants SFI-07/YI2/I1051 and 07/IN.1/I945) and by the EU-FP7 (iFOX
project). We made use of computational facilities provided by the
Trinity Centre for High Performance Computing (TCHPC) and the Irish
Centre for High-End Computing (ICHEC).
\end{acknowledgments}

\bibliography{references}

\end{document}